\documentclass[final
  ]
  {aipproc}

\layoutstyle{8x11double}

\begin{document}

\title{Modelling hybrid $\beta$\,Cephei\,/\,SPB pulsations: $\gamma$~Pegasi}


\author{T.~Zdravkov}{address={Nicolaus Copernicus Astronomical Center, ul. Bartycka 18, 00-716 Warsaw, Poland}}

\author{A.~A.~Pamyatnykh}{address={Nicolaus Copernicus Astronomical Center, ul. Bartycka 18, 00-716 Warsaw, Poland}
,altaddress={Institute of Astronomy, Russian Academy of Science, Pyatniskaya Str.48, 109017 Moscow, Russia}}

\begin{abstract}
Recent photometric and spectroscopic observations of the hybrid variable $\gamma$~Pegasi
\cite{Hetal09, H09} revealed 6 frequencies of the SPB type and 8 of the $\beta$\,Cep type pulsations.
Standard seismic models, which have been constructed with OPAL \cite{IR96} and OP \cite{S05}
opacities by fitting three frequencies (those of the radial fundamental and two dipole modes),
do not reproduce the frequency range of observed pulsations and do not fit the observed individual
frequencies with a satisfactory accuracy. We argue that better fitting can be achieved with
opacity enhancements, over the OP data, by about 20-50 percent around the opacity bumps produced
by excited ions of the iron-group elements at temperatures of about $2\times10^{5}$\,K (Z~bump) and
$2\times10^{6}$\,K (Deep~Opacity~Bump).
\end{abstract}

\maketitle


\section{Introduction}
Hybrid stars are main-sequence variables which show two
different types of oscillations: (i) low-order acoustic and gravity
modes of $\beta$\,Cephei type with periods of about $3-6$ hours, and
(ii) high-order gravity modes of the SPB type with periods of about
$1.5-3$ days. Pulsations of both types of variables are driven by the iron opacity bump located
at temperature of about $2\times10^{5}$~K (Z~bump).
In the HR diagram, the overlapping region of the $\beta\,$Cep and SPB type
variables is very sensitive to the opacity data (Figs.\,3 and 4 in \cite{P99},
see also \cite{M07a, M07b, ZP08a}), therefore the modelling of hybrid star pulsations allows to test the opacity
of the stellar matter.

It was suggested by \cite{DP08} that an opacity enhancement around the Z bump
could help solving the problems in seismic modelling of the hybrid variable $\nu$\,Eri.
We tested this possibility in \cite{ZP08b}.
There is also another iron opacity bump, referred to as Deep Opacity Bump (DOB),
at a temperature of about $2\times10^{6}$~K, which
may be responsible for the excitation of g-modes in hot massive Wolf-Rayet stars \cite{TM06}.
An opacity enhancement in the DOB temperature region (more exactly, at the base
of the solar convection zone) was suggested recently to achieve agreement between
new solar models and helioseismic inferences \cite{ChD09}.

Here, we construct seismic models of the hybrid variable $\gamma$~Pegasi,
using the OPAL~\cite{IR96} and OP~\cite{S05} data on stellar opacities
for two heavy element mixtures: GN93~\cite{GN93} and A04~\cite{A04}. We show
that some opacity enhancements around the Z bump and Deep Opacity Bump may be required to
achieve a better agreement between the theoretical frequency range of unstable modes and the
observed pulsations, and also to achieve a better fit of the computed individual frequencies
to the observed values.

\section{Standard seismic models}

Recent photometric and spectroscopic observations of $\gamma$~Pegasi
\cite{Hetal09, H09} revealed 6 frequencies of the SPB type and 8 of the $\beta$\,Cep type pulsations.
Two modes were definitively identified: the radial fundamental mode at $6.5897$~cd$^{-1}$ and the
dipole mode ${\rm g}_1$ at 6.0162~cd$^{-1}$. Preliminary seismic model from \cite{Hetal09} fits these
two modes. In our study we suggested that the observed peak at 6.9776~cd$^{-1}$ corresponds
to the acoustic dipole mode ${\rm p}_1$. Our seismic models were constructed to fit these three frequencies
by a suitable choice of stellar mass, $M$, heavy element abundance, $Z$, and effective temperature.
This method was used also in our studies of $\nu$\,Eri \cite{PHD04, DP08, ZP08b}.

The position of the seismic models built with standard opacity data is shown on Fig.\,1.
Standard models don't solve all the problems,
which are similar to those encountered in the modelling of $\nu\,$Eri \cite{DP08, ZP08b}:
{\bf (i)}~The theoretical frequency range of the unstable high-order gravity modes of lowest degrees
does not fit the observed range. In OP models only quadruple modes are unstable in the observed frequency range,
whereas the observed frequency spacings argue mainly in favour of dipole modes. In the OPAL case, SPB type
pulsations are not excited at all.
{\bf (ii)}~In models within the observational error box in the HRD
(the OPAL model and preliminary model from \cite{Hetal09}), the theoretical frequency of the dipole mode
${\rm p}_2$ is noticeably  higher than the observed value 9.109~cd$^{-1}$.
OP models fit this value much better but they are located outside the observational error box in the HRD.
{\bf (iii)}~All observed frequencies higher than 8~cd$^{-1}$ are outside
the theoretical frequency range of unstable modes.
{\bf (iv)}~Observations suggest rather low metallicity (up to $Z=0.01$), whereas one needs much
higher $Z$ values to explain the mode excitation in the whole observed frequency range.

\begin{figure}
  \includegraphics[height=.2\textheight]{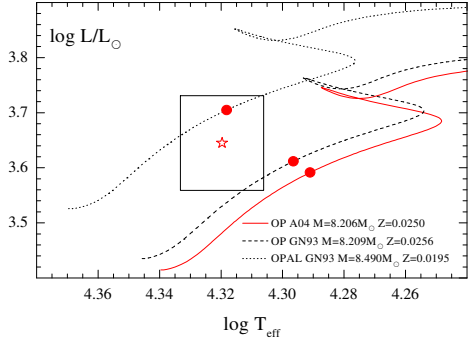}
  \caption{Influence of the choice of the opacity data (OP vs. OPAL) and the heavy element mixture
(A04 vs. GN93) on seismic model of $\gamma$~Peg. The models which fit both radial fundamental and
two dipole modes (${\rm g}_1$ and ${\rm p}_1$) are marked with filled circles. The observational error box is due
to Kurucz model atmospheres and Hipparcos parallax.}
\end{figure}

\section{Models with opacity enhancements}
\paragraph{\bf Effect of the 50\% opacity increase in the Z bump}
The Z bump at a temperature of about $2\times10^{5}$\,K is located in relatively low-density layers,
therefore the opacity modification in this region does not change
frequencies of the lowest acoustic and gravity  modes that are used to construct seismic models.
As a consequence, the position of the modified seismic model in the HR diagram is almost the same as that
of the standard model (see Fig.\,4 below).
However, the opacity enhancement results in an extension of the unstable frequency range because just the Z bump
is responsible for the driving of both SPB and $\beta$\,Cep type pulsations, see Fig.\,2. Note that
in the modified model the frequency of the third dipole mode (mode ${\rm p}_2$ at 9.1~cd$^{-1}$) is somewhat
smaller than that in the standard model and than the observed value.

\begin{figure}
  \includegraphics[height=.25\textheight]{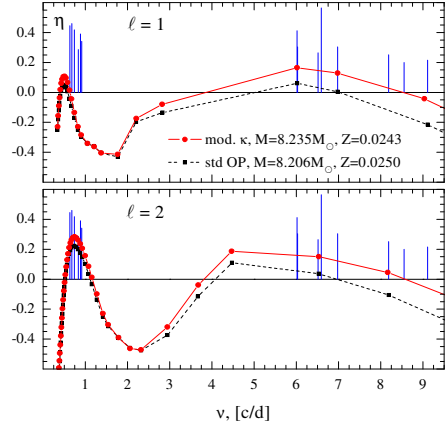}
  \caption{Normalized growth rate ($\eta>0$ for instability) as a function of frequency of dipole
and quadruple modes in seismic models of $\gamma$~Peg: the standard OP model (dashed line) and the model
with the the 50\% opacity enhancement in the Z bump region (solid line).
Vertical bars represent the observed oscillation spectrum, with amplitudes in the logarithmic scale.}
\end{figure}

\paragraph{\bf Effect of the 20\% opacity increase in the Deep Opacity Bump}
The Deep Opacity Bump at a temperature of about $2\times10^{6}$\,K is located in more dense layers
than the Z bump. Therefore, the opacity modification influences even lowest acoustic
and gravity frequencies including those used to construct the fitted seismic model.
The modified seismic model noticeably differs from the standard one and
its position in the HR diagram is different. The model is located within the observational error box
(see Fig.\,4 below). On the other hand, these deep stellar layers produce only marginal driving,
therefore the instability of the modified model is almost the same as in the standard case, see Fig.\,3.
Contrary to the Z bump case, the frequency of the third dipole mode (mode ${\rm p}_2$ at 9.1~cd$^{-1}$)
is now somewhat higher than that in the standard model and than the observed value.

\begin{figure}
  \includegraphics[height=.25\textheight]{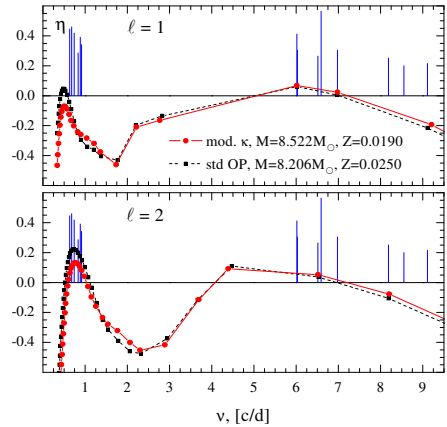}
  \caption{Normalized growth rate of the dipole and quadruple modes in seismic model with
with the 20\% opacity enhancement in the DOB region.}
\end{figure}

\paragraph{\bf Combined effect of the opacity modifications in both regions}
The modified model was constructed with an opacity enhancement (over the OP A04 data) of 50\%
in the Z bump and of 16\% in the DOB. The second value was chosen
to fit nicely the frequency of the dipole mode ${\rm p}_2$, as well as the frequencies
of the radial fundamental and two other dipole modes. The modified model is located within
the observational error box, as shown on Fig.\,4, and it has a significantly lower Z value than
the standard OP A04 model. Moreover it is in better agreement with the observational results (see \cite{Hetal09}
and references therein).

\begin{figure}
  \includegraphics[height=.2\textheight]{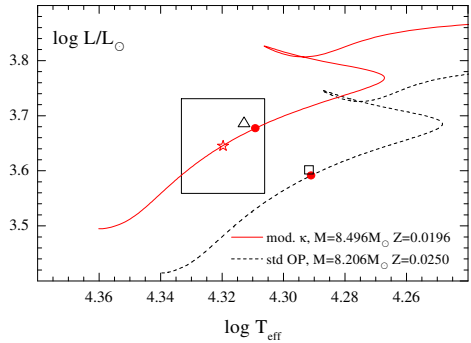}
  \caption{The standard OP A04 model and the models with enhanced opacity in the HR diagram.
Evolutionary tracks correspond to the standard seismic model and to the seismic model with
the opacity enhancement both by 50\% in the Z bump and by 16\% in the DOB.
These models are marked with filled circles. The seismic models with the
opacity enhancement only in the Z bump region (by 50\%) and only in the DOB region (by 20\%)
are marked with open square and open triangle, respectively.}
\end{figure}

\begin{figure}
  \includegraphics[height=.25\textheight]{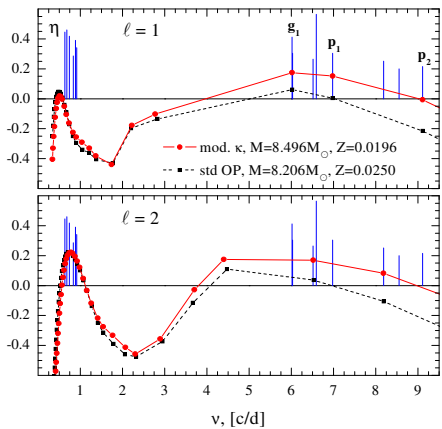}
  \caption{Normalized growth rate of the dipole and quadruple modes in the seismic model with opacity enhancements
in both iron bumps: 50\% in the Z bump region and 16\% in the DOB region.
Labels above the bars indicate mode identifications of the dipole modes used to construct the seismic model.}
\end{figure}

The oscillations modes of the $\beta$\,Cep type are unstable in the whole observed frequency range.
As in the standard case, only quadruple high-order gravity modes are unstable in the observed frequency
range of the SPB type pulsations. This model fits all observed frequencies better than other models.
Moreover, only in this model all $\beta$\,Cep type modes are unstable (the ${\rm p}_2$ mode is marginally stable).
The fitting for the presumably radial mode at 8.552~cd$^{-1}$ is not satisfactory,
but this mode has the lowest observed amplitude (see \cite{Hetal09}).

\section{Conclusions}
To explain the observed frequency spectrum of the hybrid variable $\gamma$~Pegasi, the stellar
matter seems to be still more opaque around the iron opacity bumps at temperatures of about $2\times10^{5}$~K
(Z bump) and $2\times10^{6}$~K (DOB) than given by the standard models constructed with the OPAL or OP data.
A model with an opacity enhancement (over the OP A04 data) of 50\%
in the Z bump region and of 16\% in the DOB region nicely reproduces
the $\beta$\,Cep frequencies. The model is also unstable in the high-order gravity modes of lowest degrees,
but only quadruple modes fit the observed frequency range of the SPB type pulsations, whereas the observed
frequency spacings argue mainly in favour of dipole modes \cite{Hetal09}. Note that in all our models there
is an unstable quadruple mode at 4.4\,-\,4.5~cd$^{-1}$ which is not detected in observations.


\begin{theacknowledgments}
The work was supported by the Polish MNSiW grant No. N N203 379636.
\end{theacknowledgments}

\end{document}